\documentclass[twocolumn,prd,aps,floats,nofootinbib,showpacs]{revtex4}
\usepackage{epsfig}
\usepackage{graphicx}
\usepackage{color}
\newcommand{\gtrsim}{\,\rlap{\lower3.7pt\hbox{$\mathchar\sim$}}
\raise1pt\hbox{$>$}\,}
\newcommand{\lesssim}{\,\rlap{\lower3.7pt\hbox{$\mathchar\sim$}}
\raise1pt\hbox{$<$}\,}

\newcommand{\be}{\begin{equation}}
\newcommand{\ee}{\end{equation}}
\newcommand{\bea}{\begin{eqnarray}}
\newcommand{\eea}{\end{eqnarray}}
\def\d{{\rm d}}
\def\alt{\raise0.3ex\hbox{$\;<$\kern-0.75em\raise-1.1ex\hbox{$\sim\;$}}}
\def\agt{\raise0.3ex\hbox{$\;>$\kern-0.75em\raise-1.1ex\hbox{$\sim\;$}}}
\begin{document}

\title{Revisiting cosmological bounds on  radiative neutrino lifetime}
\author{Alessandro Mirizzi}
\affiliation{Max-Planck-Institut f\"ur Physik
(Werner-Heisenberg-Institut), F\"ohringer Ring 6, 80805 M\"unchen,
Germany}

\author{Daniele Montanino}
\affiliation{Dipartimento di Fisica and Sezione INFN di Lecce,
             Via Arnesano, 73100 Lecce, Italy}

\author{Pasquale D.~Serpico}
\affiliation{Center for Particle Astrophysics, Fermi National
Accelerator Laboratory, Batavia, IL 60510-0500 USA}

\begin{abstract}
Neutrino oscillation experiments and direct bounds on absolute
masses constrain neutrino mass differences to fall into the
microwave energy range, for most of the allowed parameter space. As
a consequence of these recent phenomenological advances, older
constraints on radiative neutrino decays based on  diffuse
background radiations and assuming strongly hierarchical masses in
the eV range are now outdated. We thus derive new bounds on the
radiative neutrino lifetime using the high precision cosmic
microwave background spectral data collected by the Far Infrared Absolute Spectrophotometer instrument
on board of Cosmic Background Explorer. The lower bound on the lifetime is between a few$\times 10^{19}\,$s
and $\sim 5\times 10^{20}\,$s, depending on the neutrino mass ordering and
on the absolute  mass scale. However, due to phase space limitations,
the upper bound in terms of the effective magnetic moment mediating the
decay is not better than $\sim 10^{-8}\,$ Bohr magnetons.
We also comment about possible improvements of
these limits, by means of recent diffuse infrared photon background
data. We compare these bounds with pre-existing limits coming from
laboratory or astrophysical arguments. We emphasize the
complementarity of our results with others available in the
literature.
\end{abstract}
\pacs{13.35.Hb, 
95.30.Cq    
98.70.Vc    
98.80.-k,    
\hfill FERMILAB-PUB-07-135-A; MPP-2007-63}

\maketitle
\section{Introduction}\label{intro}
The last decade  has seen two great improvements in
astroparticle physics: the wealth of information
on neutrino physics (for a recent review  see e.g. \cite{Fogli2006})
and the impressive precision of new cosmological data (see for example
the latest WMAP team results \cite{Spergel:2006hy}). The standard cosmological
scenario predicts the existence of a diffuse background of
low-energy neutrinos, which has been often investigated in the past to
probe non-standard neutrino properties (for a few recent examples,
see \cite{Beacom:2004yd,Hannestad:2005ex,Mangano:2006ar,Chu:2006ua}).
At this stage, it is meaningful to reassert the impact
of non-standard physics in the neutrino sector on cosmological
observables, or equivalently to re-explore the constraints provided by
cosmology on exotic physics.

In this paper we revisit the bounds on
neutrino radiative lifetime coming from cosmology. Indeed, older
constraints based on the diffuse Cosmic Infrared Background (CIB)
and assuming strongly
hierarchical masses in the eV range \cite{Ress90,Biller:1998nc}
(see also \cite{Dolgov:2002wy,Raffeltbook,Raffelt:1999tx} for general reviews)
 are now outdated and strictly speaking inapplicable.
The neutrino mass splittings squared provided by oscillation
experiments and present upper bounds on the neutrino
mass scale constrain neutrino mass differences to fall in the microwave
 energy range ($E \sim 10^{-3}$~eV), in most of the allowed parameter space.
 Proper bounds
must be derived using the high precision cosmic microwave background
(CMB) data collected by the Far Infrared Absolute Spectrophotometer (FIRAS) instrument on board the Cosmic Background Explorer (COBE), which
tested the blackbody nature of the spectrum at better than 1 part in
10$^{4}$ \cite{Fixsen:1996nj,Mather:1998gm}. The high precision of
this measurement has been already exploited to constrain some new
physics scenarios, producing deformations on the CMB
spectrum~\cite{Mirizzi:2005ng,Melchiorri:2007sq}. In addition, a new
estimate of the CIB flux (although less accurate than CMB) has been recently
derived~\cite{Dole:2006de} from the SPITZER
telescope data~\cite{Werner:2004zk}. Consequently, one can also infer
updated bounds on the radiative lifetime when the unredshifted
photon energy falls in this infrared  range ($E \sim 10^{-2}$~eV),
i.e., in the limit of non-degenerate neutrino mass pattern.

Here we perform such an analysis for radiative neutrino decays,
 and compare the new  bounds
obtained with limits coming from laboratory or astrophysical arguments.
Although taken at face value these bounds are not competitive with
the most stringent ones already available, we emphasize that they probe different
combinations and/or regimes of the effective couplings describing the
electromagnetic properties of neutrinos, thus being complementary (rather then
redundant) with respect to the others available in the literature.
The plan of our work is as follows.
In Sec. \ref{raddecay} we summarize the relevant formalism, while in
Sec. \ref{data} we present the CMB data  used and the bounds obtained.
In Sec. \ref{CIB} we briefly discuss the bounds coming from the CIB flux.
In Sec. \ref{concl} we comment our results and give the conclusion.

\section{radiative neutrino decays}\label{raddecay}
Let us denote by $\nu_i$ the (active) neutrino fields respectively
of masses  $m_i$, where $i=1,2,3$. The radiative decay $\nu_i\to
\nu_j+\gamma$ can be thought of as arising from an {\it effective}
interaction Lagrangian of the form
\be {\cal L}_{\rm
int}=\frac{1}{2}\bar{\nu}^i\sigma_{\alpha\beta}(\mu_{ij}+\epsilon_{ij}\gamma_5)\nu^jF^{\alpha\beta}+\:{\rm
h.c.} \label{Eq1}\ee
where $F^{\alpha\beta}$ is the electromagnetic field tensor,
$\sigma_{\alpha\beta}=[\gamma_\alpha,\gamma_\beta]$ where
$\gamma_\mu$ are the Dirac-matrices and $[.\,,.]$ is the commutator,
$\nu_i$ is the neutrino field of mass $m_i$, and $\mu_{ij}$ and
$\epsilon_{ij}$ are the magnetic and electric transition moments
usually expressed in units of the Bohr magneton $\mu_B$. The
convention to sum over repeated indices is used. In general,
$\mu_{ij}$ and $\epsilon_{ij}$ are functions of the transferred
momentum squared $q^2$, so that constraints obtained at a different $q^2$
are independent. The radiative decay rate for a transition $i\to j$
is written
\begin{eqnarray}
\Gamma^\gamma_{ij}&=&\frac{|\mu_{ij}|^2+|\epsilon_{ij}|^2}{8\pi}\left(\frac{
m_i^2-m_j^2}{m_i}\right)^3 \nonumber \\
&\equiv&
\frac{\kappa_{ij}^{2}}{8\pi}\left(\frac{m_i^2-m_j^2}{m_i}\right)^3.\label{kappa}
\end{eqnarray}

In the following, we shall quote the bounds in terms of
$\kappa_{ij}^{2}$. We shall
assume that the radiative decay rate is very low compared with the
expansion rate of the universe; neither the cosmological evolution
or the primordial neutrino spectrum is affected by the additional
coupling we are going to introduce. A posteriori, this is known to
be an excellent approximation. For the same reason, we shall also
neglect ``multiple decays" (the daughter neutrino $\nu_j$
constitutes a negligible fraction of the original $\nu_i$
quasi-thermal population). We shall take our input data for neutrino
mass eigenstate densities from the calculation performed in
\cite{Mangano:2005cc} without any extra parameter, as non-vanishing
chemical potentials. With present data, the latter are anyway
constrained to be well below ${\cal O}$(1) \cite{Serpico:2005bc}, so
dropping this assumption would not change much our conclusions.

From simple kinematical considerations it follows that in a decay
$\nu_i\to \nu_j+\gamma$ from a state of mass $m_i$ into one of mass
$m_j < m_i$, the photon  in the rest frame of the decaying
neutrinos is thus monochromatic (two-body decay), with an energy
\be \varepsilon_{ij} = \frac{m_i^2-m_j^2}{2\, m_i}\,. \ee

At present, the neutrino mass spectrum is constrained
by  the well-known values of the two squared mass splittings for the
atmospheric ($\Delta m_{H}^2$) and the solar  ($\Delta
m_{L}^2$) neutrino problems. We take their
best-fit values and $2\sigma$ ranges from \cite{Fogli2006}:
\begin{eqnarray}
\Delta m_{L}^2 &=& 7.92\, (1\pm 0.09)\times 10^{-5}\mathrm{\ eV}^2\ ,\\
\Delta m_{H}^2 &=& 2.6\,(1^{+0.14}_{-0.15}) \times
10^{-3}\mathrm{\ eV}^2\ .
\end{eqnarray}
The remaining unknowns in the neutrino spectrum are the absolute mass
scale (equivalently, the mass of the lightest eigenstate $m_1$) and
the mass hierarchy. Namely, in normal hierarchy (NH)   the mass pattern
would be
\begin{eqnarray}
&m_1& , \nonumber \\
 &m_2&=\sqrt{m_1^2+\Delta m_{L}^2} \,\ , \nonumber \\
&m_3&=\sqrt{m_1^2+\Delta
m_{L}^2+\Delta m_{H}^2} \,\ ;
\end{eqnarray}
while in inverted hierarchy (IH) one would have
\begin{eqnarray}
 &m_1& , \nonumber \\
 &m_2&=\sqrt{m_1^2+\Delta m_{H}^2} \,\ , \nonumber \\
&m_3&=\sqrt{m_1^2+\Delta m_{L}^2+\Delta m_{H}^2} \,\ .
\end{eqnarray}

In the limiting case of normal hierarchy and $m_1 = 0$, the lightest
neutrino for which a decay is possible has a mass $m_2 \simeq 9\times
10^{-3}\,$eV and is thus non-relativistic for most of the universe
lifetime, namely in the redshift range $z\alt 50$. We can thus
safely work in the approximation of all neutrinos decaying
effectively at rest. In this limit, we can also neglect the momentum
distribution of the neutrino spectra. The formalism which would
allow one to generalize our results to the momentum-dependent case has
been developed in \cite{Masso:1999wj}, which we address the interested reader for further details. However, the corrections are small, of the order of powers
of the neutrino temperature to mass ratios, and also vanishing in the
limit of very long lifetimes.

We shall discuss the limits on $\kappa_{ij}^{2}$ as a function of
$m_1$ and for the two cases NH and IH. We shall vary the mass scale
in $0\lesssim m_1\lesssim 2\,$eV as allowed by the Mainz experiment
on the ${}^3$H beta decay endpoint \cite{Kraus:2004zw}. In this
respect, we shall be conservative: If neutrinos are Majorana
particles the more stringent bound from $0\nu\beta\beta$ searches
apply, with an effective mass bound $m_{\beta\beta}<0.81\,$eV
\cite{Fogli:2006yq}. Structure formation, combined with
other cosmological data, also constrains $\sum_i m_i$
\cite{Spergel:2006hy}. Present cosmological bounds span the range
$\sum_i m_i\alt 0.2-2\,$eV,
\cite{Spergel:2006hy,Lesgourgues:2006nd,Fogli:2006yq,Kristiansen:2006ky,Melchiorri:2007cd,Hannestad:2007dd}
depending on the data sets used and priors assumed. An upper limit
of $\sum_i m_i\sim\,$0.6 eV (i.e. $m_1\simeq 0.2\,$eV) is often
considered robust, and we shall report it for illustrative purposes.
Yet, one may circumvent the $0\nu\beta\beta$ bound (e.g. with a
Dirac neutrino) and significantly relax the most stringent
cosmological bounds (for example with a conservative combination of
cosmological data sets and priors or with an exotic dark energy
sector), so in the following we shall present our results up to the
value $m_1=2\,$eV.

Let $F_E$ be the present energy flux of photons with present energy
$E$ produced by neutrino decay. The differential energy flux $\varphi_E$ (energy
flux $F_E$ per unit energy and solid angle) is related to the differential
number flux $\varphi_n$ (the particle flux $F_n$ per unit energy and solid angle) at present by
\begin{equation}
\varphi_E\equiv\frac{\d^2F_E}{\d E\, \d\Omega} = E\, \frac{\d^2F_n}{\d E\,
\d\Omega}\equiv E\,\varphi_n,
\end{equation}
and it can be shown that, if the lifetime $\tau_i$ of the
neutrino of mass $m_i$ is much greater than the universe lifetime
it holds  \cite{Masso:1999wj} \footnote{
Note that $\tau_i$ may be much shorter than the radiative lifetime,
which in most exotic models is dominated by invisible
decays~\cite{Hannestad:2005ex,Fogli:2004gy,Serpico:2007pt}.
In the present work we are neglecting these cases.}
\begin{equation} \label{restspectr}
\varphi_E = \frac{\Gamma^{\gamma}_{32}}{4\pi}
\frac{n_{3}}{H(z_{32})}+
\frac{\Gamma^{\gamma}_{31}}{4\pi}\frac{n_{3}}{H(z_{31})}+
\frac{\Gamma^{\gamma}_{21}}{4\pi}
\frac{n_{2}}{H(z_{21})}\,,
\end{equation}
where $n_i \simeq 113$~cm$^{-3}$ is the present number density of the $i-$th neutrino {\it in
absence of decay}, the Hubble function is (assuming, for simplicity,
a flat cosmology)
$H(z)=H_0\sqrt{\Omega_M(1+z)^3+\Omega_\Lambda}$, $H_0\simeq 73$~km~s$^{-1}$~Mpc$^{-1}$
 being the
present Hubble expansion rate, and $\Omega_M \simeq 0.26$ and
$\Omega_\Lambda \simeq 0.74$
respectively the matter and the cosmological constant energy density
relative to the critical one.  The dependence on energy enters
implicitly via the quantities $0\leq z_{ij}=\varepsilon_{ij}/E-1$.

In practice, to a very good approximation one can write a general equation of the kind
\begin{equation} \label{restspectr2}
\varphi_E = \frac{\Gamma^{\gamma}_{H}}{4\pi}
\frac{n_{H}}{H(z_H)}+\frac{\Gamma^{\gamma}_{L}}{4\pi}
\frac{n_{L}}{H(z_L)}\,,
\end{equation}
where, however, the meaning of the factors however depends on the hierarchy.
In NH,
in the first two terms of the sum in Eq.~(\ref{restspectr}) it holds
$z_{32}\simeq z_{31}\equiv z_H$, and one can identify
 $z_L=z_{21}$, $\Gamma^{\gamma}_L=\Gamma^{\gamma}_{21}$,
$\Gamma^{\gamma}_H=\Gamma^{\gamma}_{31}+\Gamma^{\gamma}_{32}$. In
IH, it is the last two terms of the sum in Eq. (\ref{restspectr})
which have $z_{31}\simeq z_{21}\equiv z_H$, and using $n_2\simeq n_3$
one can identify $z_L=z_{32}$, $\Gamma^{\gamma}_L=\Gamma^{\gamma}_{32}$,
$\Gamma^{\gamma}_H\equiv\Gamma^{\gamma}_{31}+\Gamma^{\gamma}_{21}$.
In both cases, we shall therefore express our bounds in terms of
$\kappa_{L,H}^2$ keeping in mind their slightly different meaning
for the two cases of NH and IH.

\begin{figure}[t]
\centering
\epsfig{figure=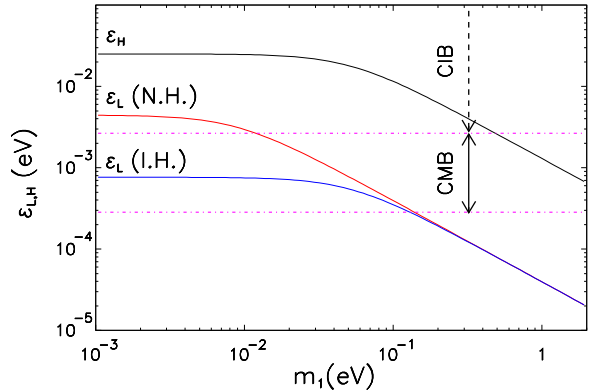,width =1.\columnwidth,angle=0}
 \caption{\label{fig1} Unredshifted photon energy $\varepsilon$
   from decaying
  neutrinos [Eq.~(3)] as a function of the lightest neutrino mass
  eigenstate $m_1$, for the two neutrino mass splittings (L,H)
 in normal and inverted hierarchy. (See text for details)
The horizontal band represents the energy range of the CMB spectrum
 measured by FIRAS~\cite{Fixsen:1996nj}. The CIB energy range is also shown.}
\end{figure}

In Fig.~1 we represent the unredshifted photon energy $\varepsilon_{ij}$
from decaying neutrinos [Eq.~(3)] as a function of the lightest neutrino
mass eigenstate $m_1$ in the case of normal and inverted hierarchy,
where the meaning of $\varepsilon_{L,H}$ is clear from the previous discussion.
We also indicate by an horizontal band
the energy range of the CMB spectrum
($2.84\times10^{-4}\,{\rm eV}\leq E \leq 2.65\times 10^{-3}\,$eV)
 measured by FIRAS~\cite{Fixsen:1996nj}.
We also show the CIB range in the energy band above the
FIRAS range up to (conventionally) 0.15~eV~\cite{Dole:2006de}. For
$m_1\alt 0.5$~eV, the photon energy $\varepsilon_H$ falls in the CIB range.

 For photons emitted at $z=0$ in the FIRAS range, the effect of radiative decays is most
prominent and results in a feature on the CMB spectrum.
 Actually even if photons
are emitted at higher energy the effect is still strong, since
photons emitted at a redshift of a few enter the FIRAS spectrum
because of cosmological redshift; it is easy to check that one has
thus some sensitivity to $\kappa_{H}$ in the whole range for $m_1$.
However, as we will see, for $m_1 \lesssim 0.1$~eV a stronger (but less robust) limit can be
obtained using directly the CIB data.

On the other hand, for $m_1\agt 0.14$ eV the photons corresponding
to the smaller splitting are falling in the radio band, below the frequency
range probed by COBE, where measurements are more uncertain
 and thus one has no sensitivity to $\kappa_L$ and
the corresponding bound disappears.

\section{The CMB bound}\label{data}

To constrain the neutrino electromagnetic decay we use the COBE/FIRAS data for
the experimentally measured CMB spectrum, corrected for
foregrounds~\cite{Fixsen:1996nj}. Note that the new calibration of
FIRAS~\cite{Mather:1998gm} is within the old errors and would not
change any of our conclusions. The $N = 43$ data points $\Phi^{\rm
exp}_i$ at different energies $E_i$ are obtained by summing the
best-fit blackbody spectrum to the residuals reported in
Ref.~\cite{Fixsen:1996nj}.  The experimental errors $\sigma_i$ and
the  correlation indices $\rho_{ij}$ between different energies are
also available.  In the presence of neutrino decay, the original
radiance (energy flux per unit of solid angle) of the ``theoretical blackbody'' at temperature $T$
\begin{equation}
\label{planck1} \Phi^0(E,T) = \frac{E^3}{4 \pi^3} \big[
\exp (E/T )-1 \big]^{-1}
\end{equation}
would gain an additional term so that the intensity becomes
\begin{eqnarray}
\label{planck2}
\Phi^0(E,T)&\to&\Phi(E,T,\kappa_{L,H}^2, m_1) \nonumber \\
&=& \Phi^0(E,T) +\varphi_E (\kappa_{L,H}^2, m_1) .
\end{eqnarray}

\begin{figure}[t]
\vspace*{-5mm}
\centering
\epsfig{figure=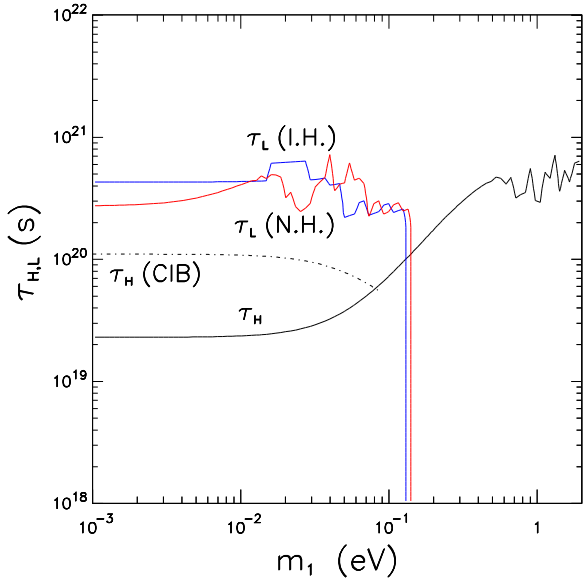,width =1.\columnwidth,angle=0}
 \caption{\label{fig3} Bounds on $\tau_{H}$ and $\tau_{L}$ vs. $m_1$, for
the two cases of NH and IH.
 The regions below the solid curves are excluded at 95 \% C.L.
 The curves for $\tau_H$ coincide in the two
cases, although the definition of $\tau_H$ is different (see text).
The dot-dashed line represents the limit on $\tau_H$ obtained from
cosmic infrared background.}
\end{figure}

We then build the reduced chi-squared function
\begin{equation}
\chi_\nu^2(T,\kappa_{L,H}^2, m_1)=\frac{1}{{N}-1} \sum_{i,j=1}^{N} {\Delta
\Phi_i} (\sigma^2)^{-1}_{ij} {\Delta \Phi_j} \,,
\end{equation}
where
\begin{equation}
\Delta \Phi_i = \Phi^{\rm exp}_i-\Phi(E_i,T)
\end{equation}
is the $i$-th residual, and
\begin{equation}
\sigma^2_{ij}= \rho_{ij} \sigma_{i} \sigma_{j}
\end{equation}
is the covariance matrix. In principle, the parameter $T$ entering
initially in $\Phi^0(E,T)$ needs not to be fixed at the standard
value $T_0=2.725\pm0.002$~K \cite{Mather:1998gm}, which is the best
fit of the ``distorted'' spectrum eventually observed now. The
initial $T$ before a significant fraction of neutrinos decays should
thus be a free parameter, to be determined in the minimization
procedure. Practically, however, the distortion introduced by the
neutrino decay spectrum is such highly non-thermal that a change in
$T$ can not be accomodated for any significant neutrino lifetime:
the constraints obtained fixing $T$ are basically the same.

\begin{figure}[t]
\vspace*{-5mm}
\centering
\epsfig{figure=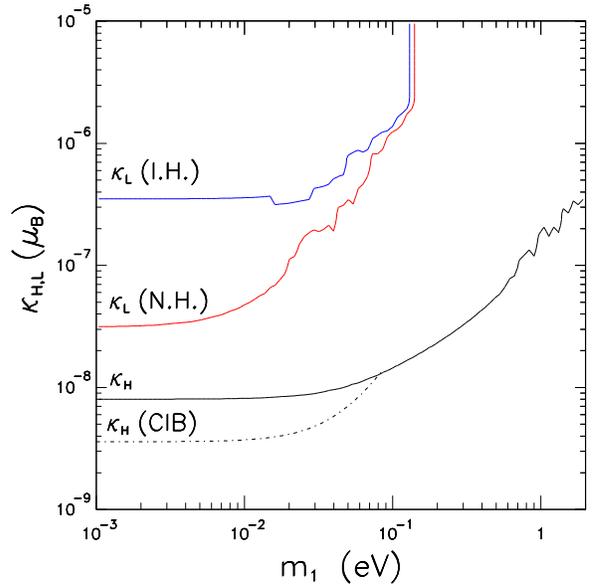,width =1.\columnwidth,angle=0}
 \caption{\label{fig2} Bounds on $\kappa_{H}$ and $\kappa_{L}$ vs. $m_1$, for
the two cases of NH and IH.
 The regions above the  solid curves are excluded at 95 \% C.L.
 The curves for $\kappa_H$ coincide in the two
cases, although the definition of $\kappa_H$ is different (see
text). The dot-dashed line represents the limit on $\kappa_H$
obtained from cosmic infrared background.}
\end{figure}

 Our results
are reported in Fig.~2, where we represent the  exclusion plot in
the plane $\tau_{L,H}\equiv (\Gamma^\gamma_{H,L})^{-1}$ vs. $m_{1}$,
where the regions below the solid curves are excluded at 95 \% C.L.
For small values of $m_1$ the most stringent limit is $\tau_L
\gtrsim 4 \times 10^{20}$~s in  IH (slightly better
than in NH case), while the bound on $\tau_H$ is about an order of
magnitude smaller, say $\tau_H \gtrsim 2 \times 10^{19}$~s, since for
low $m_1$ only photons produced by $H$ decays at a redshift of few
are in FIRAS range. On the contrary, for $m_1 \gtrsim 0.14$~eV, the
bound on $\tau_L$ disappears, while the bound on $\tau_H$ becomes
more stringent, being $\tau_H \gtrsim 5 \times 10^{20}$~s.
Note that the ``fuzzy'' behaviour of the bounds  is due to the sharp edge of the photon
spectrum at $E=\varepsilon_{H,L}$: when the photon energy  embeds a new FIRAS bin,
the $\chi^2$ function has a sharp discontinuity.

In Fig.~3 we translate the plot of Fig.~2 in an exclusion plot in the plane
$m_1$ vs. $\kappa_{L,H}$. Here the factor $(\delta m_{ij}^2/m_i)^3$
maps in a non trivial way the bounds in terms of
$\kappa_{L,H}$. The regions above the solid curves are excluded at 95
\% C.L. For the NH case, $\kappa_L\alt 3\times 10^{-8}\,\mu_B$,
while in the IH case, $\kappa_L\alt 3\times 10^{-7}\,\mu_B$. In
agreement with our previous considerations, the bound on $\kappa_L$
disappears for $m_1\agt 0.14\,$eV. On the contrary, the bound for
$\kappa_H$ is always present, and it corresponds to $\kappa_H\alt 8
\times 10^{-9}\,\mu_B$ apart for the degenerate region, where it
degrades down to $10^{-7}\,\mu_B$ or even more. Note also that 
typical cosmological upper bounds would already exclude the extreme 
degenerate case.

\section{The CIB bound}\label{CIB}

The Cosmic Infrared Background (CIB) is mainly the relic emission of
the formation and evolution of the galaxies of all types at
wavelengths larger than a few microns. The spectrum of the CIB is
peaked around $\sim 100$~$\mu$m ($E \sim 1.2\times 10^{-2}$~eV),
thus just is in the energy range $\varepsilon_H$ of photon from
radiative $\nu$ decays, for $m_1<0.1$~eV. Recently, a new estimate
of the CIB flux  has been established  using the Spitzer Observatory
data~\cite{Dole:2006de}. The  measured CIB flux  is $\Phi_{\rm
CIB}\sim 24$~nW~m$^{-2}$~sr$^{-1}$. Using this number we can obtain
a rough bound on $\tau_H$ (and hence on $\kappa_H$) simply requiring
that the total energy flux of the photons coming from $\nu$ decay
does not exceed the CIB flux:
\begin{equation}
\label{eq:CIB_lim}
\int_{E_{\rm min}}^{\varepsilon_H}\varphi_E\, dE<\Phi_{\rm CIB} \,\ ,
\end{equation}
where we consider as lower limit of the CIB range  the upper value
of the FIRAS range, i.e. $E_{\rm min}=2.65\times 10^{-3}$~eV. The
bounds of $\tau_H$ and $\kappa_H$ obtained from
Eq.~(\ref{eq:CIB_lim}) are shown respectively in Figs.~2 and 3 by
the dot-dashed line~\footnote{ For NH \ and $m_1\lesssim
10^{-2}$~eV also $\varepsilon_L$ falls marginally in the CIB range.
However, we have explicitly checked that the FIRAS constraint on
$\varepsilon_L$ is always stronger.}. Although these bounds are
stronger than those obtained by the FIRAS data in the same range of
$m_1$, we emphasize that they should be considered only as
indicative. In fact, the CIB flux have still strong uncertainties
and the precise spectral shape is essentially unknown (a factor
$\sim 3$ of uncertainty should be accounted~\cite{Dole:2006de}).

It is interesting to comment that, if it turns out that $m_1\alt
0.1\,$eV, an improvement on the bound on $\tau_{\rm H}$ will clearly
take advantage of a better measurement of the CIB flux and a more
detailed knowledge of the astrophysical sources contributing to it.
Conversely,  $m_1\agt 0.1\,$eV would imply a significant degree of
neutrino clustering in large dark matter halos, the larger the mass
the stronger the clustering \cite{Singh:2002de,Ringwald:2004np}. In
turn, the expectation of overdensities would motivate analyses in
the microwave sky toward specific targets (like nearby galaxies or
galaxy clusters), thus taking full advantage of the spectral feature
expected from neutrino decay and looking for an angular-dependent
enhancement over the CMB background. In spirit, this would be
similar to what performed in the X-ray band when searching for
signatures of sterile neutrino or axion decays (see e.g.
\cite{Watson:2006qb,Riemer-Sorensen:2007qw}).
 Although a detailed treatment of these issues
goes beyond the purpose of this work, it is worth noting that which
one is the regime to consider will be basically answered by the
KATRIN experiment on tritium beta decay \cite{Osipowicz:2001sq}.

\section{Discussion and Conclusions}\label{concl}

In this paper, we have revisited the bounds on the neutrino radiative
lifetime coming from cosmology, deriving updated constraints from
the high precision CMB spectrum data collected by the FIRAS
instrument on board of COBE. We also compare these bounds with those
obtained (by a back-of-the-envelope calculation) using the measurement of the flux of
the Cosmic Infrared Background, which  although sometimes
overrides those coming from the CMB data, should be considered
only as qualitative.

 Previous cosmological bounds were either
derived in a pre-COBE era \cite{Ress90} or there were assumed higher masses
and/or a strongly hierarchical mass spectrum, thus using the
infrared background to derive the constraints \cite{Biller:1998nc}.
This has motivate us to re-evaluate the bounds within the presently
allowed range of parameters suggested by neutrino oscillation
physics and tritium endpoint experiments. Since it is customary to
parameterize the neutrino electromagnetic decay via an effective
operator of the kind reported in Eq.~(\ref{Eq1}), it makes sense to
translate the bounds (which actually are  on the lifetime) into
bounds on the parameters $\kappa_{H,L}$ [see
Eqs.~(\ref{kappa},\ref{restspectr2})]. These constraints are not
better than $\kappa\alt 10^{-8}\,\mu_B$ which at first sight do not
appear competitive with astrophysical limits neither with most of
the laboratory bounds \cite{Yao:2006px}. Nevertheless, every
experimental measure and every cosmological and astrophysical
constraint has its own systematic uncertainties and its own
recognized or un-recognized loop-holes. In this sense  it is
certainly important to use
 many different approaches to constrain fundamental neutrino properties.

In particular, the cosmological bound is based on the appearance of
the daughter photons, and thus is very direct (modulo the underlying
cosmological assumptions). Therefore the bounds on $\tau_{H,L}$ are
completely independent from the underlying model that mediates the
neutrino decay. Each decay model must face with our direct bounds on
$\tau_{H,L}$ wich are the strongest attainable with direct livetime
measurements. Moreover, it also probes the energy scale  $q^2\alt
10^{-3}\,{\rm eV}^2$, unaccessible to both laboratory experiments
and stellar arguments. Other constraints have different features.
Neutrino electromagnetic couplings are actually tightly constrained
by energy-loss arguments in stars, in particular via the plasmon
process $\gamma^{*}\to \bar{\nu}\nu$, to be $\kappa_{ij}\alt 3\times
10^{-12}\,\mu_B$ \cite{Raffelt:1989xu,Raffelt:1998xu}. These are
indirect limits, which strictly speaking apply to $q^2\agt {\rm
keV}^2$. Laboratory bounds are obtained via elastic $\nu\,e$
scattering, where the scattered neutrino is not observed, and are at
most at the level of $10^{-10}\,\mu_B$ for the electron flavor
\cite{Yao:2006px,Pakvasa:1999ta}. The combinations of matrix
elements $\epsilon_{ij}$ and $\mu_{ij}$ that are constrained by
various experiments depend on the initial neutrino flavor and on its
propagation between source and detector. Cancellations may occur in
exotic cases \cite{Raffeltbook}, and in any case the energy scale
probed is $q^2\agt {\rm MeV}^2$.
 An additional motivation
for independent checks is that models with a strong dependence of
$\kappa_{ij}(q^2)$ have been proposed \cite{Frere:1996gb}. Our bound
seems to exclude extreme runnings of the effective couplings. Vice
versa, if one assumes that the elements $\epsilon_{i,j}$ and
$\mu_{ij}$ are quasi energy independent, laboratory and
astrophysical arguments exclude any possibility of
phenomenologically interesting radiative decays in cosmology; this
is a priori surprising, given that cosmology involves the longest
time intervals available and the lowest boosting factors for
neutrinos. On the other hand, note that the cosmological bound may
be easily violated by invoking a neutrino invisible decay much
faster than the universe lifetime. It is intriguing to notice that,
despite the impossibility to test this ``nightmare case" in
laboratory, cosmology may probe to some extent those scenarios,
potentially with important consequences for particle physics as well
\cite{Serpico:2007pt,Fogli:2004gy}.

\section*{Acknowledgments}
We thank J. Beacom and G. Raffelt for useful comments on the
manuscript. In Germany the work of A.M. is supported by an Alexander
von Humboldt fellowship grant. In Italy the work of D.M. is
supported in part by the Italian ``Istituto Nazionale di Fisica
Nucleare'' (INFN) and by the ``Ministero dell'Istruzione,
Universit\`a e Ricerca'' (MIUR) through the ``Astroparticle
Physics'' research project. P.S. is supported by the US Department
of Energy and by NASA grant NAG5-10842.


\begin{thebibliography}{100}
\bibitem{Fogli2006}
  G.~L.~Fogli, E.~Lisi, A.~Marrone and A.~Palazzo,
  ``Global analysis of three-flavor neutrino masses and mixings,''
  Prog.\ Part.\ Nucl.\ Phys.\  {\bf 57}, 742 (2006)
  [hep-ph/0506083].

\bibitem{Spergel:2006hy}
  D.~N.~Spergel {\it et al.}  [WMAP Collaboration],
  ``Wilkinson Microwave Anisotropy Probe (WMAP) three year results:
  Implications for cosmology,''
  astro-ph/0603449.

\bibitem{Beacom:2004yd}
  J.~F.~Beacom, N.~F.~Bell and S.~Dodelson,
  ``Neutrinoless universe,''
  Phys.\ Rev.\ Lett.\  {\bf 93}, 121302 (2004)
  [astro-ph/0404585].

\bibitem{Hannestad:2005ex}
  S.~Hannestad and G.~Raffelt,
  ``Constraining invisible neutrino decays with the cosmic microwave
  background,''
  Phys.\ Rev.\  D {\bf 72}, 103514 (2005)
  [hep-ph/0509278].

\bibitem{Mangano:2006ar}
  G.~Mangano  {\it et. al.},
  ``Effects of non-standard neutrino electron interactions on relic  neutrino
  decoupling,''
  Nucl.\ Phys.\ B {\bf 756}, 100 (2006)
  [hep-ph/0607267].

\bibitem{Chu:2006ua}
  Y.~Z.~Chu and M.~Cirelli,
 ``Sterile neutrinos, lepton asymmetries, primordial elements: How much of each?,''
  Phys.\ Rev.\  D  {\bf 74}, 085015 (2006)
  [astro-ph/0608206].

\bibitem{Ress90}
M.T. Ressell and M.S. Turner,
``The grand unified photon spectrum: a coherent view
of the diffuse extragalactic background radiation,''
 Comm. Astrophys. {\bf 14}, 323 (1990).

\bibitem{Biller:1998nc}
  S.~D.~Biller {\it et al.},
  ``New limits to the IR background: Bounds on radiative neutrino decay and  on
  VMO contributions to the dark matter problem,''
  Phys.\ Rev.\ Lett.\  {\bf 80}, 2992 (1998)
  [astro-ph/9802234].

  \bibitem{Dolgov:2002wy}
  A.~D.~Dolgov,
  ``Neutrinos in cosmology,''
  Phys.\ Rept.\  {\bf 370}, 333 (2002)
  [hep-ph/0202122].

\bibitem{Raffeltbook}
  G.~G.~Raffelt,
  {\it Stars as Laboratories for Fundamental Phys\-ics},
  (Univ.\ of Chicago Press, 1996).

\bibitem{Raffelt:1999tx}
  G.~G.~Raffelt,
  ``Particle physics from stars,''
  Ann.\ Rev.\ Nucl.\ Part.\ Sci.\  {\bf 49}, 163 (1999)
  [hep-ph/9903472].

\bibitem{Fixsen:1996nj}
  D.~J.~Fixsen, {\it et al.},
  ``The cosmic microwave background spectrum from the full
  COBE/FIRAS data set,''
  Astrophys.\ J.\ {\bf 473}, 576 (1996)
  [astro-ph/9605054].

\bibitem{Mather:1998gm}
  J.~C.~Mather, D.~J.~Fixsen, R.~A.~Shafer, C.~Mosier
  and D.~T.~Wilkinson,
  ``Calibrator design for the COBE far infrared absolute
  spectrophotometer (FIRAS),''
  Astrophys.\ J.\  {\bf 512}, 511 (1999)
  [astro-ph/9810373].

\bibitem{Mirizzi:2005ng}
  A.~Mirizzi, G.~G.~Raffelt and P.~D.~Serpico,
  ``Photon axion conversion as a mechanism for supernova dimming: Limits  from
  CMB spectral distortion,''
  Phys.\ Rev.\  D {\bf 72}, 023501 (2005)
  [astro-ph/0506078].

\bibitem{Melchiorri:2007sq}
  A.~Melchiorri, A.~Polosa and A.~Strumia,
  ``New bounds on millicharged particles from cosmology,''
 	Phys.\ Lett.\ B, {\bf 650} 416 (2007)

\bibitem{Dole:2006de}
  H.~Dole {\it et al.},
  ``The Cosmic Infrared Background Resolved by Spitzer. Contributions of
  Mid-Infrared Galaxies to the Far-Infrared Background,''
  Astron.\. Astrophys.\ {\bf 451}, 417 (2006).

\bibitem{Werner:2004zk}
  M.~W.~Werner {\it et al.},
  ``The Spitzer Space Telescope Mission,''
  Astrophys.\ J.\ Suppl.\  {\bf 154} 1 (2004)
  [astro-ph/0406223].

\bibitem{Mangano:2005cc}
  G.~Mangano {\it et al.},
  ``Relic neutrino decoupling including flavour oscillations,''
  Nucl.\ Phys.\  B {\bf 729}, 221 (2005)
  [hep-ph/0506164].

\bibitem{Serpico:2005bc}
  P.~D.~Serpico and G.~G.~Raffelt,
  ``Lepton asymmetry and primordial nucleosynthesis in the era of precision
  cosmology,''
  Phys.\ Rev.\  D {\bf 71}, 127301 (2005)
  [astro-ph/0506162].

\bibitem{Masso:1999wj}
  E.~Mass\`o and R.~Toldr\`a,
  ``Photon spectrum produced by the late decay of a cosmic neutrino
  background,''
  Phys.\ Rev.\  D {\bf 60}, 083503 (1999)
  [astro-ph/9903397].

  \bibitem{Kraus:2004zw}
  C.~Kraus {\it et al.},
  ``Final results from phase II of the Mainz neutrino mass search in  tritium
  beta decay,''
  Eur.\ Phys.\ J.\  C {\bf 40}, 447 (2005)
  [hep-ex/0412056].

\bibitem{Fogli:2006yq}
  G.~L.~Fogli {\it et al.},
  ``Observables sensitive to absolute neutrino masses: A reappraisal after
  WMAP-3y and first MINOS results,''
  Phys.\ Rev.\  D {\bf 75}, 053001 (2007)
  [hep-ph/0608060].

\bibitem{Lesgourgues:2006nd}
  J.~Lesgourgues and S.~Pastor,
  ``Massive neutrinos and cosmology,''
  Phys.\ Rept.\  {\bf 429}, 307 (2006)
  [arXiv:astro-ph/0603494].

\bibitem{Kristiansen:2006ky}
  J.~R.~Kristiansen, O.~Elgaroy and H.~Dahle,
``Using the cluster mass function from weak lensing to constrain
neutrino masses,''
  arXiv:astro-ph/0611761.

\bibitem{Melchiorri:2007cd}
  A.~Melchiorri, O.~Mena and A.~Slosar,
``An improved cosmological bound on the thermal axion mass,''
  arXiv:0705.2695 [astro-ph].


\bibitem{Hannestad:2007dd}
  S.~Hannestad, A.~Mirizzi, G.~G.~Raffelt and Y.~Y.~Y.~Wong,
``Cosmological constraints on neutrino plus axion hot dark matter,''
  arXiv:0706.4198 [astro-ph].

\bibitem{Fogli:2004gy}
  G.~L.~Fogli, E.~Lisi, A.~Mirizzi and D.~Montanino,
  ``Three-generation flavor transitions and decays of supernova relic
  neutrinos,''
  Phys.\ Rev.\  D {\bf 70}, 013001 (2004)
  [hep-ph/0401227].

\bibitem{Serpico:2007pt}
  P.~D.~Serpico,
  ``Cosmological neutrino mass detection: The best probe of neutrino
  lifetime,''
  Phys.\ Rev.\ Lett.\  {\bf 98}, 171301 (2007)
  [astro-ph/0701699].


\bibitem{Singh:2002de}
  S.~Singh and C.~P.~Ma,
``Neutrino clustering in cold dark matter halos: Implications for
ultra high energy cosmic rays,''
  Phys.\ Rev.\  D {\bf 67}, 023506 (2003)
  [astro-ph/0208419].

\bibitem{Ringwald:2004np}
  A.~Ringwald and Y.~Y.~Y.~Wong,
``Gravitational clustering of relic neutrinos and implications for
their detection,''
  JCAP {\bf 0412}, 005 (2004)
  [hep-ph/0408241].

\bibitem{Watson:2006qb}
  C.~R.~Watson, J.~F.~Beacom, H.~Yuksel and T.~P.~Walker,
``Direct X-ray constraints on sterile neutrino warm dark matter,''
  Phys.\ Rev.\  D {\bf 74}, 033009 (2006)
  [astro-ph/0605424].


\bibitem{Riemer-Sorensen:2007qw}
  S.~Riemer-Sorensen, {\it et al.},
  ``Searching for decaying axion-like dark matter from clusters of galaxies,''
  astro-ph/0703342.



\bibitem{Osipowicz:2001sq}
  A.~Osipowicz {\it et al.}  [KATRIN Collaboration],
``KATRIN: A next generation tritium beta decay experiment with
sub-eV sensitivity for the electron neutrino mass,'' hep-ex/0109033.


\bibitem{Yao:2006px}
  W.~M.~Yao {\it et al.}  [Particle Data Group],
  ``Review of particle physics,''
  J.\ Phys.\ G {\bf 33}, 1 (2006).

  \bibitem{Raffelt:1989xu}
  G.~G.~Raffelt,
  ``Core mass at the Helium flash from
 observations and a new bound
on neutrino electromagnetic properties,''
  Astrophys.\ J.\  {\bf 365}, 559 (1990).

\bibitem{Raffelt:1998xu}
  G.~G.~Raffelt,
  ``Comment on neutrino radiative decay limits from the infrared  background,''
  Phys.\ Rev.\ Lett.\  {\bf 81}, 4020 (1998)
  [astro-ph/9808299].


\bibitem{Pakvasa:1999ta}
  S.~Pakvasa,
  ``Do neutrinos decay?,''
  AIP Conf.\ Proc.\  {\bf 542}, 99 (2000)
  [hep-ph/0004077].

\bibitem{Frere:1996gb}
  J.~M.~Frere, R.~B.~Nevzorov and M.~I.~Vysotsky,
  ``Stimulated neutrino conversion and bounds on neutrino magnetic moments,''
  Phys.\ Lett.\  B {\bf 394}, 127 (1997)
  [hep-ph/9608266].



    \end{thebibliography}
  \end{document}